# Cumulative Polarization Coexisting with Conductivity at Interfacial Ferroelectrics

Swarup Deb[1#], Wei Cao[2#], Noam Raab[1], Kenji Watanabe[3], Takashi Taniguchi[4], Moshe Goldstein[1], Leeor Kronik[5], Michael Urbakh[2], Oded Hod[2], Moshe Ben Shalom[1*]

[1] School of Physics and Astronomy, Tel Aviv University, Tel Aviv, Israel

[2] Department of Physical Chemistry, School of Chemistry, The Raymond and Beverly Sackler Faculty of Exact Sciences and The Sackler Center for Computational Molecular and Materials Science, Tel Aviv University, Tel Aviv, Israel.

[3] Research Center for Functional Materials, National Institute for Material Science, 1-1 Namiki, Tsukuba 305-0044, Japan

[4] International Center for Materials Nanoarchitectonics, National Institute for Material Science, 1-1 Namiki, Tsukuba 305-0044, Japan

[5] Department of Molecular Chemistry and Materials Science, Weizmann Institute of Science, Rehovoth, Israel

[#] These authors contributed equally

* moshebs@tauex.tau.ac.il

Ferroelectricity in atomically thin bilayer structures has been recently predicted[1] and measured[2–4] in two-dimensional (2D) materials with hexagonal non-centrosymmetric unit-cells. Interestingly, the crystal symmetry translates lateral shifts between parallel 2D layers to a change of sign in their out-of-plane electric polarization, a mechanism referred to as "Slide-Tronics"[4]. These observations, however, have been restricted to switching between only two polarization states under low charge carrier densities[5–12], strongly limiting the practical application of the revealed phenomena[13]. To overcome these issues, one needs to explore the nature of the polarization that arises in multi-layered van der Waals (vdW) stacks, how it is governed by intra- and inter-layer charge redistribution, and to which extent it survives the introduction of mobile charge carriers, all of which are presently unknown[14]. To explore these questions, we conduct surface potential measurements of parallel $WSe_2$ and $MoS_2$ multi-layers with aligned and anti-aligned configurations of the polar interfaces. We find evenly spaced, nearly decoupled potential steps, indicating highly confined interfacial electric fields, which provide means to design multi-state "ladder ferroelectrics". Furthermore, we find that the internal polarization remains significant upon electrostatic doping of a mobile charge carrier density as high as $10^{13}$ cm$^{-2}$, with substantial in-plane conductivity. Using first-principles calculations based on density functional theory (DFT), we trace the extra charge redistribution in real and momentum space and identify an eventual doping-induced depolarization mechanism.

Electronic devices based on 2D materials utilize electric fields that add or remove mobile charge carriers to control the material conductivity. If these materials also possess switchable electric dipoles, they could serve as the basis of future high-density memory technology[13–16]. Therefore, a visionary goal is to design 2D materials that simultaneously exhibit non-volatile memory and rapid logic response, where internal charge dipoles and free electrons coexist[17,18]. Such coexistence, however, is typically unfavored since free charge carriers tend to screen dipole formation and cooperative orientation[19–23]. In this respect, the recent discovery of interfacial ferroelectricity by interlayer vdW sliding suggests an approach to overcome the above difficulty by exploiting a partition between in- and out-of-plane phenomena. Here, in-plane conductivity is afforded by a conducting 2D electron/hole gas, whereas out-of-plane switchable polarization emerges from intrinsic symmetry breaking at the interface that can be controlled by interlayer sliding. This inherent anisotropy of layered ferroelectrics should further distinguish them from thin ferroelectric films, which are highly susceptible to depolarizing fields at the surface or interface where the polarization terminates[24]. Specifically, at the 2D limit, the polarization magnitude rarely scales with the system thickness and can switch between two local states only[25–27].

To demonstrate and explore the potential of interfacial ferroelectrics, we study devices made of two or three layers of transition metal dichalcogenides (TMDs) that are artificially stacked in a parallel lattice orientation and encapsulated by thin flakes of non-polar hexagonal boron nitride (*h*-BN), placed atop a graphite or gold metallic electrode (Fig. 1a, inset). We measure the surface potential at room temperature, ~10 nm above the surface with an atomic force microscope operated in a side-band Kelvin probe mode (KPFM) (SI. S1)[2,4]. The obtained potential map is presented in Fig. 1a, showing a triangular domain landscape of various polarization values, separated by thin domain walls that naturally form due to a slight twist angle between the flakes. For a sufficiently small global twist, these domain walls accommodate a shear displacement of precisely one interatomic spacing, allowing for perfect high symmetry AB/BA stacking (zero twist) within the triangular domains[28–30]. Measuring the potential profile across domain walls, we find five distinct polarization values (Fig. 1b). The potential profile measured in the top-left triangular region in Fig. 1a (red line) exhibits two polarization states, consistent with previous reports on non-centrosymmetric bilayer TMDs incorporating a single polar interface[1,9,11]. This indicates regions of the $WSe_2$ trilayer sample where only one of the two interfaces is active, namely exhibiting finite polarization due to non-centrosymmetric stacking and in-plane atomic relaxation. Notably, the corresponding profile measured at the central region (dashed blue line) shows three polarization states separated by potential steps of $\Delta V_{KP}$~110 mV (Fig. 1b). The potential of the intermediate step is the average of the two potential values corresponding to a single active-interface trilayer, suggesting two oppositely polarized (↑↓) interfaces within a trilayer $WSe_2$ region, as in the case of mirror-symmetric Bernal stacking (ABA) of $WSe_2$. This interpretation is further supported by the fact that the potential differences between the three polarization states equal those measured at the single-active interface regions, which translates to absolute polarization values twice as large as those measured for the bilayer system. This, in turn, can be achieved if the two interfaces have parallel polarizations (↑↑, ↓↓), which is the case for the rhombohedral ABC and CBA stacking configurations.

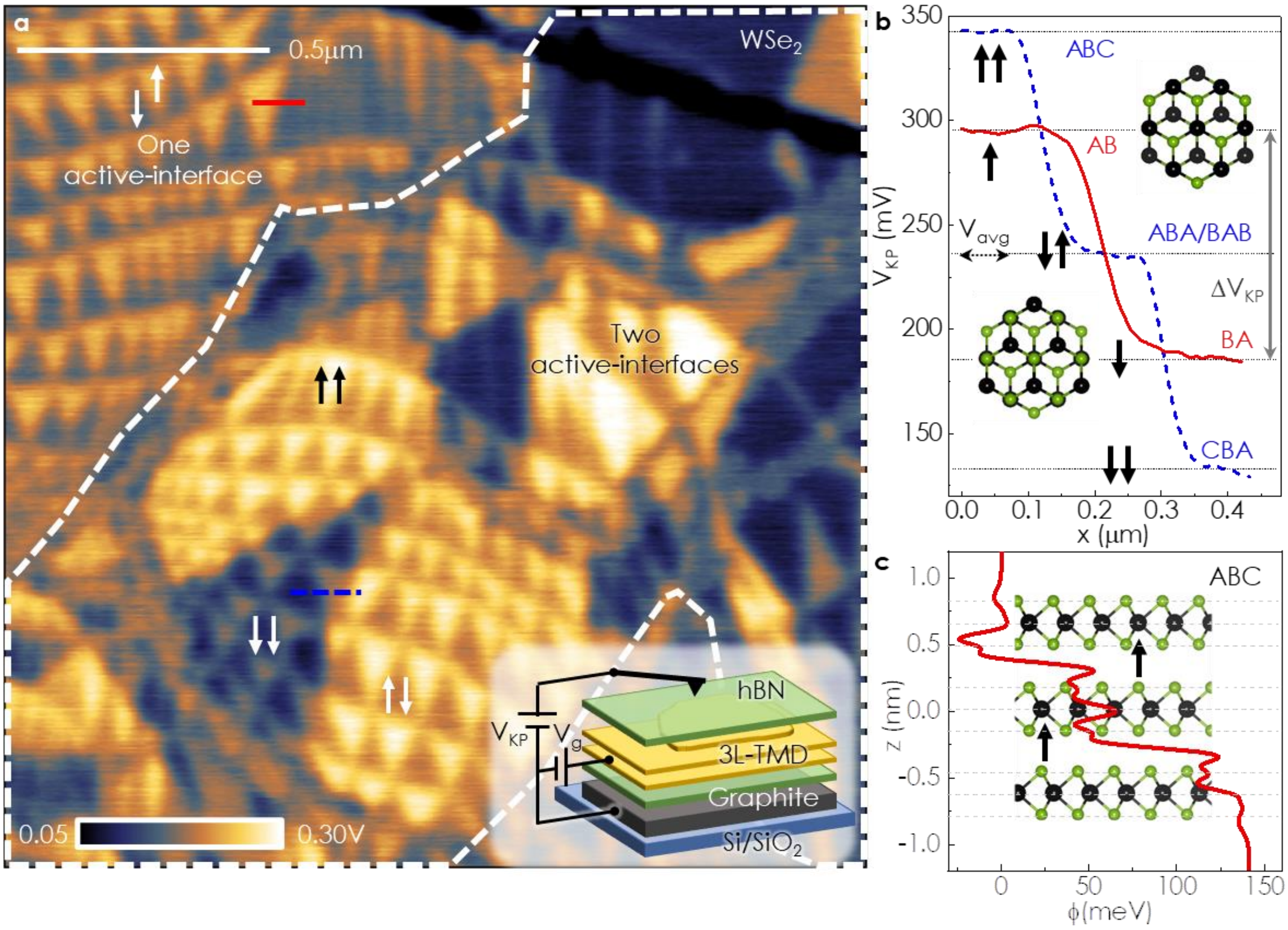

**Fig. 1| Multiple polarization states in artificially-stacked trilayers. a.** Surface potential map at the $WSe_2$ tri-layer surface. The dashed white lines mark borders between regions consisting of single and two active interfaces. Arrows denote the out-of-plane polarization orientation in five domains with different stacking configurations. Inset, schematic illustration of the Kelvin probe force microscopy setup including a tri-layer TMDs structure (yellow), encapsulated between *h*-BN layers (green) that are placed on a graphitic or gold (black) gate ($V_g$) electrode lying atop a silicon oxide substrate (blue). **b.** Typical line cuts of the lateral potential drop across walls separating large domains (extracted from the map in Fig. S1f) of single (solid red) and two (dashed blue) active interface regions. Line cuts crossing domain walls that separate smaller domains are illustrated by the red and dashed blue lines in panel **a**. The stacking configurations at the corresponding interfaces (shown schematically for the bilayer) are marked aside each potential step, with the corresponding interface polarizations marked by black arrows. **c.** Calculated laterally-averaged vertical potential profile along an ABC stacked trilayer with co-oriented (black arrows) interfacial polarization.

The experimental evidence presented above indicates that the polarization should be localized at the interfaces between layers, suggesting that adjacent interfaces are only weakly coupled and, therefore, a cumulative polarization effect in layered stacks is obtained. This is further supported by the comparable coverage of the ABC and ABA domains in the map, demonstrating no significant energetic stability preference between the two stacking configurations[29,30]. The former domain, which have two aligned polar interfaces, thus exhibit a similar stacking energy to that of the latter domains, which include two anti-aligned polar interfaces. We note, however, that for larger domains small coverage differences are observed, indicating weak Ferro-like coupling that favors a co-aligned polar ABC configuration (as discussed in SI.S2). Fig. 1c presents DFT-computed potential profiles for the ABC stacked $WSe_2$ trilayer. The potential drop, $\Delta\phi$, calculated between points far above and below the layered system is in good agreement with the measured potential drop ($\Delta V_{KP} = 2\Delta\phi$) and its step-like shape[31,32] further emphasizes the interfacial confinement of the polarization, and hence the weak coupling between adjacent polarized interfaces.

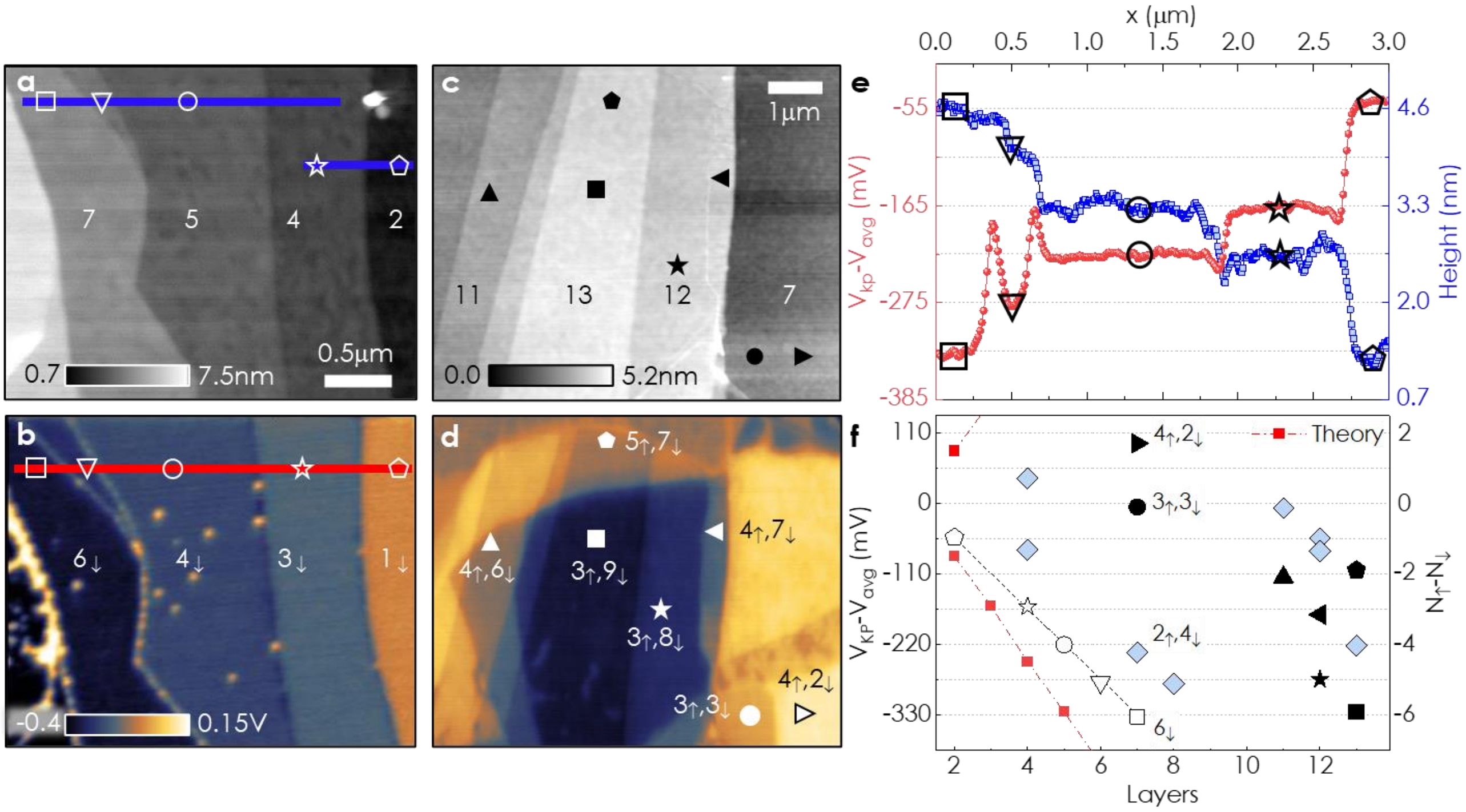

**Fig. 2| Multi-polarization states in naturally-grown 3R $MoS_2$. a-d.** Topography (a,c) and surface potential (b,d) maps of two typical flakes composed of 2-7, and 7-13, respectively. The potential is measured relative to the value above an ABA stacked tri-layer region ($V_{avg}$ in Fig.1b). **e.** Line cuts, as marked in a,b, showing the flake thickness (blue squares, left axis) and surface potential (red circles, right axis). The horizontal grids show evenly-spaced steps. **f.** Surface potential values and excess number of active interfaces ($N_{\uparrow}$ - $N_{\downarrow}$) above different positions (as marked in a-d) versus the number of layers at each point. The dashed black line connects points of the fully co-aligned polar interfaces, where the symbols correspond to those appearing in panels a, b and e. Other points (with symbols corresponding to those appearing in panels c and d) show fixed, evenly spaces values corresponding to multi-polarization configurations. For example, the four values measured above 7 layers with 6 active interfaces correspond to the $6_{\downarrow}$, $5_{\downarrow}1_{\uparrow}$, $3_{\downarrow}3_{\uparrow}$, and $2_{\downarrow}4_{\uparrow}$ combinations. Calculated maximal polarization values are indicated by the full red squares.

To demonstrate the emergence of multi-polarization states beyond tri-layered stacks, we measured the potential at the surface of $MoS_2$ crystals, which are naturally grown in the 3R ABC stacking configuration. The addition of layers with aligned polarization results in an essentially linear increase of the total polarization with stack thickness (Fig. 2a,b,e), confirming the cumulative interfacial effect. Remarkably, some region of the flakes reveal various potential values indicative of multiple interfacial polarization configurations (Fig. 2c,d) of aligned and anti-aligned polarized interfaces. For a given number of layers, regions of different stacking and polarization are spatially separated by local domain walls, whose crossing results in evenly spaced potential steps. The specific value of the measured potential above each region is determined by the difference between upward ($N_\uparrow$) and downward ($N_\downarrow$) polarization pointing active interfaces, which is dictated by the local stacking configuration and can be extracted from the measured local potential (Fig.2f). In case of 7 layers, for example, with N=6 interfaces the system can exhibit N+1 polarization values.[33] Therefore, by a relative shift of each pair of adjacent layers one could, in principle,[4] increase or decrease the surface potential in a sequential ladder of polarization values.

The interface-localized nature of the polarization paves the way to an even more unusual effect, namely its coexistence with in-plane conductance through the individual layers. We set out to explore this intriguing possibility by introducing external gate electrodes to induce free charge carriers in polarized MoS2 or WSe2 bilayers. In Fig. 3a-c, we map the potential surface of the same spatial MoS2 region, while applying several fixed gate biases Vg. Already upon the application of a relatively small bias, one notices a conductance response (Fig. S1c), as well as an improvement in the map quality (Fig. 3a,b), indicating that, indeed, the gate bias affects the carrier density in the bilayers. The application of a larger gate bias leads to domain wall sliding and a reversible polarization orientation switching, as reported here and previously for bilayer systems4 (see Fig. S2 for additional examples). The results of $\Delta V_{KP}$ measurements under different gate biases are presented in Fig. 3d for MoS2 (red stars) and WSe2 (blue triangles and circles) bilayers. The displacement field D and the carrier density n for each gate bias value are extracted from the change in the average surface potential, $V_{avg}$, between the two domains (marked in Fig. 1b). This procedure (see SI.S3 for more details) is insensitive to effects such as quantum capacitance or Schottky barriers (the latter prevented us from attaining hole doping in MoS2). Notably, the polarization in both materials is sustained up to the highest experimentally accessible charge density of $n \approx 10^{13}\ \mathrm{cm}^{-2}$. A reduction of 25-50% in the polarization, however, is observed at $\mathrm{n} \approx \pm 3 \times 10^{12}\ \mathrm{cm}^{-2}$. These findings are in qualitative agreement with DFT calculations, also shown in Fig. 3d (solid red and dashed blue lines for MoS2 and WSe2, respectively), in which doping is introduced by the inclusion of “pseudoatoms” with fractional nuclear charges, allowing the introduction of excess free charge carriers without violating sample neutrality and without distorting the underlying band-structure, see SI.S5)34. The experimental polarization is known to provide a lower bound to the true polarization35, owing to limitations of the local potential measurements under external bias and screening effects due to contaminants accumulating atop the surface at large carrier densities (see SI.S3). This explains the underestimation of the experimental measurements with respect to the calculated values. Notably, a qualitative difference between the calculated MoS2 and WSe2 polarization response to doping is observed, where the former exhibits a weaker response to hole doping than to electron doping, whereas the latter exhibits an opposite trend.

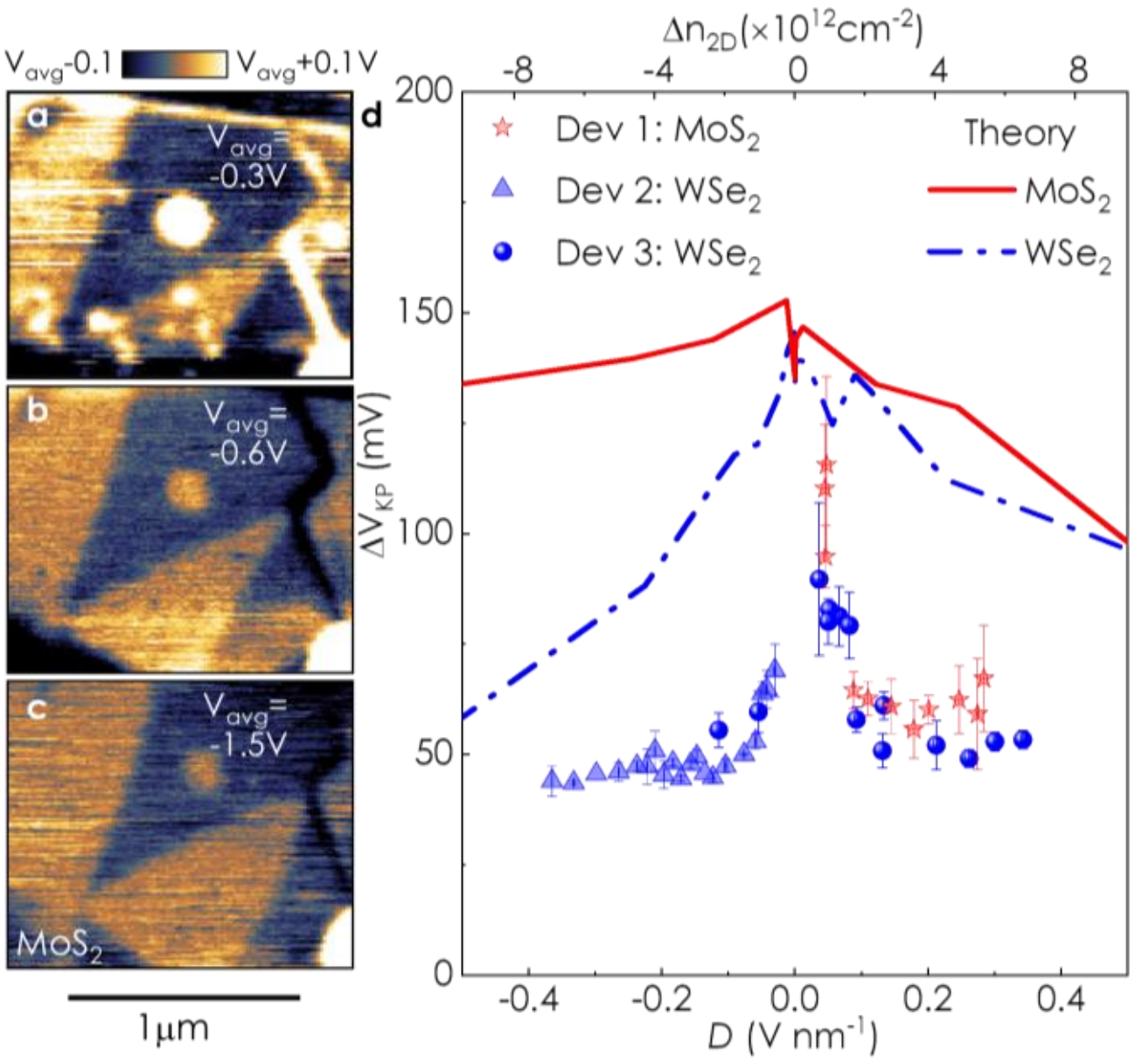

**Fig. 3| Effect of gate bias on the polarization. a-c.** Surface potential maps obtained for bilayer $MoS_2$ under different gate biases. The center of the scale bar in each map is set to the corresponding average potential, $V_{avg}$, as indicated in the respective panel. **d.** Potential drop $\Delta V_{KP}$ across domains of opposite polarization (see Fig. 1b), as a function of the external displacement field D and the corresponding 2D carrier density (lower/upper horizontal axis, see Methods section). Data from one $MoS_2$ sample (red stars) and two different $WSe_2$ samples (blue triangles and circles) are compared to the calculated $2\Delta\emptyset$ values (solid red, dashed blue lines, respectively).

To rationalize these results, we further analyze the effect of doping on the charge density distribution and its relation to the observed depolarization. Fig. 4a,b present the calculated laterally averaged excess electron charge density profiles, $\rho_0^{ex}(z)$, for the undoped $MoS_2$ (a) and $WSe_2$ (b) bilayers, where $\rho_0^{ex}(z)$ is defined as the difference between the density of the bilayer and the superposition of the densities of the corresponding undoped infinitely separated layers (dashed black lines). The excess density features a similar prominent asymmetric contribution at the interface between the two layers for both the $MoS_2$ and $WSe_2$ bilayer systems, which is the origin of the interface dipole shown in Fig. 1c (see SI.S4 for more details). Doping-induced excess charge density variations, $\Delta\rho^{ex}(z)$, are represented by colored lines for different hole densities. With increasing doping density, excess charge accumulates primarily within the layers at the transition metal plane. To analyze the asymmetry of the doping induced excess charge, which is responsible for depolarization, we plot in Fig. 4c,d the antisymmetric part of $\Delta\rho^{ex}(z)$, defined as $\Delta\rho^{ex}(z) - \Delta\rho^{ex}(-z)$, where $z = 0$ is set at the interlayer region center. For both $MoS_2$ and $WSe_2$, the asymmetric part of the excess charge shows two contributions, one at the interface and the other within the layers. We find, however, two important differences in the doping response of $\Delta\rho^{ex}(z)$ in the two materials: (i) At a given hole doping density the overall charge distribution asymmetry is larger for $WSe_2$ (compare, for example, the red curves in panels c and d of fig. 4); (ii) When integrating over the layer region excluding the interface ($z \gtrsim 0.15$nm) the asymmetric contribution of $MoS_2$ largely averages out, whereas that of $WSe_2$ does not. Note that this contribution has a stronger depolarization effect due to its larger distance from the interface. Due to both factors, depolarization is expected to commence at a much lower hole doping value in $WSe_2$ than $MoS_2$ (see Fig. S8 for comparison).

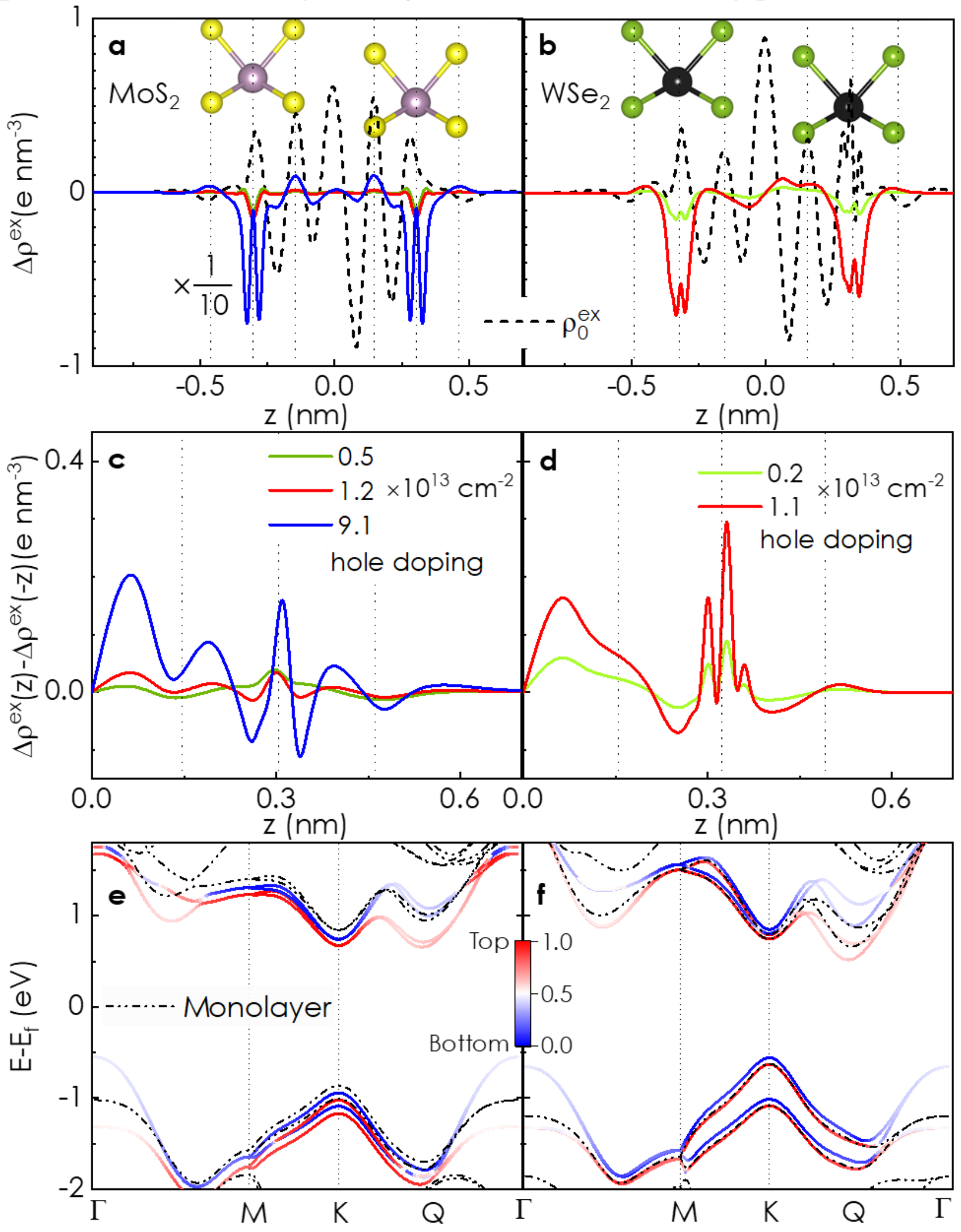

**Fig. 4| Excess charge distribution calculations for polar bilayers of $MoS_2$ (left panels) and $WSe_2$ (right panels). (a,b)** Laterally averaged excess carrier density profiles of the undoped bilayers, $\rho_0^{ex}(z)$ (dashed black line), defined as the density of the bilayer stack after subtracting the superposition charge density of two isolated layers. Doping-induced variations in the excess carrier density, $\Delta\rho^{ex}(z)$, for different hole densities are represented by the colored lines (see legends in panels c and d). All $\Delta\rho^{ex}(z)$ values for $MoS_2$ are divided by 10 for clarity of the representation. **(c,d)** Doping induced antisymmetric part of the charge carrier density variation profiles defined as $\Delta\rho^{ex}(z) - \Delta\rho^{ex}(-z)$, where $z = 0$ is set at the bilayer center for several doping densities. **(e,f)** Band structures of the undoped bilayer systems, colored according to the projection of the corresponding crystal momentum state on the top (red) and bottom (blue) layers (see text). K momentum states split into layer-specific bands (prominent red\blue) that resemble the dashed black bands of a monolayer, only separated by $\sim \frac{e\Delta V_{KP}}{2}$. A larger splitting due to strong interlayer-delocalization (designated by nearly white bands) appears for valence $\Gamma$ and conduction $Q$ states.

Finally, to explain the computationally predicted asymmetry between the polarization response to electron and hole doping, we plot in Fig. 4e,f the band structures of the two interfaces, colored according to the relative contribution of the two layers to each $k$-dependent state, $\phi_k(x,y,z)$. To this end, we evaluate its projection on the top layer as $P_k^{Top} = \int_{-\infty}^{\infty} \mathrm{d}x \int_{-\infty}^{\infty} \mathrm{d}y \int_0^{\infty} \mathrm{d}z |\phi_k(x,y,z)|^2$, with the corresponding projection on the bottom layer given by $P_k^{Bot} = 1 - P_k^{Top}$. This analysis clearly shows that the valence band states at the Γ point, $\Gamma_{VB}$, which split considerably upon the formation of the bilayer structure, are delocalized over both layers. Similar behavior is found for the conduction band states at the Q point (positioned at the midpoint between the Γ and K points of the Brillouin zone), $Q_{CB}$, at the bottom of the conduction band[31,32]. Conversely, the corresponding $K$-point states are localized on either of the layers[36] and split to a much smaller extent (comparable to $\Delta V_{KP}/2$), mainly due to the emergent interface dipole (more details in SI.S6). Therefore, changes in the occupation of the latter states will have a significant effect on the polarization. Fig. 4e shows that for $MoS_2$ the $K_{CB}$ states are encountered earlier upon raising the Fermi level (electron doping), whereas $\Gamma_{VB}$ states are encountered earlier upon lowering the Fermi level (hole doping). This explains the trend observed in Fig. 3d (solid red line), namely that depolarization commences at lower electron doping in $MoS_2$. Fig. 4f predicts an opposite behavior for $WSe_2$, owing to a much larger spin-orbit coupling induced upshift of the layer polarized $K_{VB}$ states and downshift of the layer delocalized $Q_{CB}$ states, which explains the trend shown by the dashed blue line in Fig. 3d.

To conclude, we have demonstrated stacked 2D layers that support robust interfacial polarization which features three unique characteristics: (i) it supports polarization as high as ~0.5 pC/m per interface (S1.d); (ii) it exhibits distinct and switchable multiple polarization configurations; and (iii) it sustains charge carrier densities up to $10^{14}$ $cm^{-2}$, a computational prediction that is confirmed by our current experiments for charge carrier densities as high as $10^{13}$ $cm^{-2}$. The coexistence of polarization and conductivity is attributed to the interfacial localization of the polarization and the excess charge carriers delocalization on both layers, which inhibits strong depolarization fields. Notably, the measured polarization and density values are nearly ten times larger than those found for non-hexagonal TMDs to date [5,7,8,10,37,12] and may support rich correlated electronic phases[38–40]. The cumulative distinct multi-polar ladder of states reported here thus paves the way to bottom-up construction of 3D multi-ferroic structures out of well-defined 2D building blocks in a controllable, position- and orientation-specific manner.

## Acknowledgments

We thank A. Cerreta (Park Systems) for AFM support, and N. Ravid for laboratory support. K.W. and T.T. acknowledge support from JSPS KAKENHI (Grant Numbers 19H05790, 20H00354 and 21H05233). M.G. has been supported by the Israel Science Foundation (ISF) and the Directorate for Defense Research and Development (DDR&D) grant No. 3427/21 and by the US-Israel Binational Science Foundation (BSF) Grant No. 2020072. L.K. thanks the Aryeh and Mintzi Katzman Professorial Chair and the Helen and Martin Kimmel Award for Innovative Investigation. M.U. acknowledges the financial support of the Israel Science Foundation, Grant No. 1141/18 and the binational program of the National Science Foundation of China and Israel Science Foundation, Grant No. 3191/19. O.H. is grateful for the generous financial support of the Israel Science Foundation under grant no. 1586/17, the Heineman Chair in Physical Chemistry, and the Naomi Foundation for generous financial support via the 2017 Kadar Award. M.B.S. acknowledges funding by the European Research Council (ERC) under the European Union's Horizon 2020 research and innovation programme (grant agreement no.852925), and the Israel Science Foundation under grants no. 1652/18, and 3623/21. O.H. and M.B.S. acknowledge the Center for Nanoscience and Nanotechnology of Tel Aviv University.

# Supplementary Information

## S1. Materials and methods

### a) <u>Device fabrication</u>

*h*-BN flakes of various thicknesses were exfoliated onto a Si/$SiO_2$ substrate. $MoS_2$ and $WSe_2$, obtained from HQ Graphene, were exfoliated onto polydimethylsiloxane (PDMS). Large single-layer flakes (~20 μm or more) of transition metal dichalcogenides (TMDs) were identified using optical contrast. *h*-BN flakes were picked up from the substrate and placed on few-layered graphene or predesigned gold electrodes (Fig. S1a). Subsequently, parallel bilayers of TMDs were prepared on the *h*-BN surface using the 'tear and stack' technique. In this process, a fragment of a chosen TMD flake is first stamped on *h*-BN, followed by successive stacking of the remaining flake on it. The entire stack is then encapsulated with another *h*-BN flake. The bottom graphene or gold substrate acts as a reference electrode for Kelvin probe force microscopy (KPFM) measurements and as a gate electrode. In the tri-layer measurements (without doping), the stack was placed directly on the conducting electrode without the bottom *h*-BN.

The samples studied in Fig. 2, with $MoS_2$ multi-layered systems beyond tri-layer stacks, were obtained by exfoliating the 3R single crystal (purchased from HQ graphene) to a bare $SiO_2$ surface and connecting the flakes to a metal electrode.

| **Device Name** | **Electrode** | ***h*-BN spacer thickness (nm)** |
|---|---|---|
| Dev 1: $MoS_2$ | Au | 6.3 |
| Dev 2: $WSe_2$ | Few layer graphene | 12 |
| Dev 3: $WSe_2$ | Au | 5.1 |

Table S1. Details of reported devices.

## b) AFM measurements

Topography and KPFM measurements were acquired simultaneously, using Park System NX10 AFM in non-contact scanning mode. The electrostatic signal was measured at a side-band frequency using a built-in lock-in amplifier. We used PointProbe Plus Electrostatic Force Microscopy (PPP-EFM) *n*-doped tips with a conductive coating. The mechanical resonance frequency of the tips was ~75 kHz and the force constant was 3 N/m. The cantilever oscillated mechanically with an amplitude ranging from 20 to 5 nm. In several experiments, the average height above the surface, *h,* was controlled via a two-pass measurement. The first pass records the topography, whereas in the second pass the tip follows the same scan line with a predefined lift (typically 4-5 nm) and measures the KPFM signal. The cantilever was excited with an AC voltage to perform KPFM measurements, with an amplitude of 1.5-4 V and a frequency of 2-4 kHz. In the closed-loop measurements, the DC voltage was controlled by a bias servo to obtain the surface potential. Images were acquired using the Park SmartScan software and the data analyzed with Gwyddion program.

Obtaining a reliable quantitative assessment of the absolute surface potential is challenging and requires delicate calibration processes[1–3], including the use of different tips, substrates, *h*-BN thicknesses, height of the tip above the surface, and amplitude/frequency (side-band) operation mode (see Ref. 4 of the main text). We note that the surface potential was found to be independent of the h-BN stack thickness in the range of 1-10 nm, as long as the dimensions of the domain were > 300x300 nm2. Hence, we focus our analysis on large domains, where the potential away from the domain wall saturates. Importantly, the analyses and conclusions in the present work rely on the relative change of the surface potential between two adjacent and oppositely polarized domains (see Fig. 1b), or on the relative change in the gate voltage (Fig. 3d), rather than on absolute values. The relative changes are considerably less prone to calibration uncertainties.

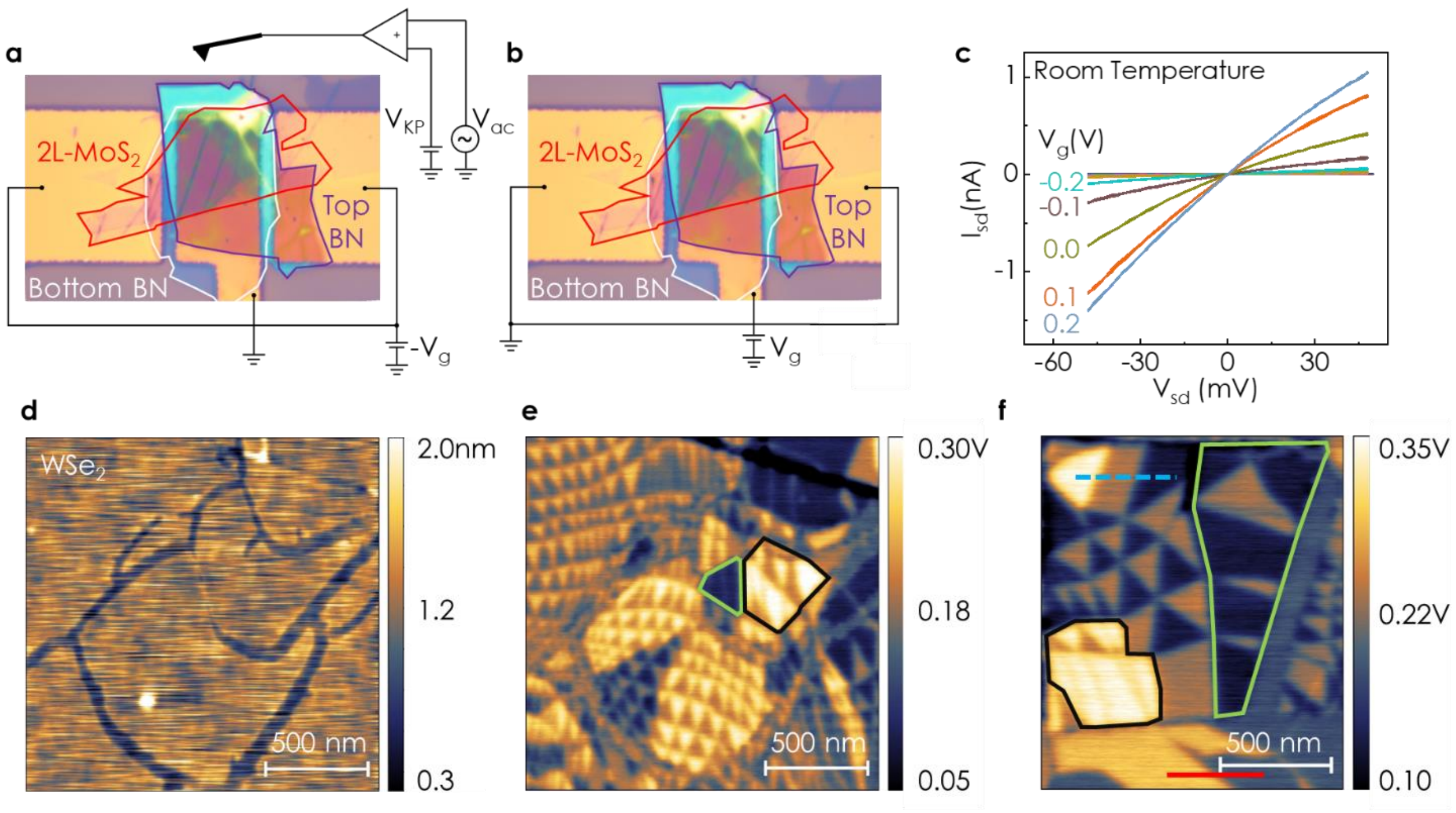

*Figure S1. Further device characterization. (a,b) Optical microscope image of a typical device. Two circuit configurations of gating and KPFM measurement are illustrated. (c) The electric current along the $MoS_2$ bilayers is measured versus the source-drain bias ($V_{sd}$) and plotted for several fixed gate voltages ($V_g$) at room temperature. (d,e) Simultaneous measurement of topography and KPFM maps for a tri-layer $WSe_2$ structure (also shown in Fig.1 in the main text). The dark regions in the topography map indicate a crack in one of the layers, separating regions of trilayers with one and two active interfaces. (e,f) Examples of ferroelectric-like coupling between the two active interfaces in large-area domains. The ↑↓ (ABA, neutral color, 0.2V) domains cover a smaller area than the ↓↓ (CBA, dark, 0.02V) domains (circumscribed in green). The same is found for ↑↑ (ABC, bright, 0.3V) regions marked in black. The solid red and dashed blue lines in (f) mark the line cut position of the data shown in the main-text, Fig. 1b.*

## c) <u>Triangular Domain Formation</u>

The domain walls exhibit an intriguing geometrical structure that emerges due to the slight twist angle between the flakes and affects the entire atomically reconstructed triangular lattice. The shear boundary conditions of the artificial stacking experiments result in a shear deformation of precisely one planar interatomic spacing in each domain wall[4] (for WSe2, each domain contributes a shear of $3.3/\sqrt{3}$ Å ). In fact, for a given global twist angle, there is a fixed number of crossing points between domain walls that is determined by the moiré superstructure topology. In our previous study (Ref. 4), we modeled the triangular superlattice structure using dedicated intra- and inter-layer force fields.

The relaxed geometry suggested that ~10 nm away from the domain wall and beyond, the stacking is perfect with zero inter-layer twists or strain, in agreement with experiment. The topological number of domain walls that end at domain wall crossings is directly related to the global twist angle in a given section by dividing the number of domain walls within the section and can be obtained by the number of interatomic spacings along the section side. For example, the scale bar in figure 1a is 0.5 μm long (~ 2,500 interatomic spacings), and it crosses ten domain walls (a total shear of 10 interatomic spacings), indicating an angle of 10/2,500 radians.

**d) <u>2D polarization calculation</u>**

The 2D polarization per interface is extracted directly from the measured $\Delta V_{\mathrm{KP}}$ of a single active interface via the relation $P_{2D} = \frac{\epsilon_0 \Delta V_{\mathrm{KP}}}{2}$. With $\epsilon_0 = 8.854 \times 10^{-12}\ \mathrm{Fm}^{-1}$ and $\Delta V_{\mathrm{KP}} \approx 120\ \mathrm{mV}$, we get $P_{\mathrm{2D}} \approx 0.53\ \mathrm{pCm}^{-1}$. This result is in agreement with the calculated value of $P_{2D} = \int_{-\infty}^{\infty} \rho_0^{ex}(z) z dz$, with $\rho_0^{ex}(z)$ being represented by the dashed black line in Fig. 4a,b.

## S2. Finite Ferroelectric-like coupling of two active interfaces

While the internal electric fields are mostly confined to the interfacial volume, as discussed in the main text, we find indications of finite coupling between adjacent regions of two active interfaces structures. This is achieved by comparing the average area coverage of co-aligned (ABC/CBA) and anti-aligned (ABA) domains (see Fig. S1e,f). The higher adhesion phase naturally expands on the expense of other stacking configurations[5]. Indeed, we find that regions in the map of large area domains (marked in black or green) show a clear preference for ABC or CBA stackings with ↑↑ or ↓↓ (bright or dark) polarization, respectively, at the expense of the anti-aligned ABA and BAB domains (with neutral-color) of ↑↓ or ↓↑ polarization, respectively. A close look at two active interface regions with smaller domains (outside the marks) also shows a reduced area of the neutral domains even away from the physical edge of the layers (although the area difference here is minor). Recently, we reported a similar behavior in a single active interface system of parallel *h*-BN bilayers, where domain wall sliding in response to an externally applied electric field promoted larger domains that align with the external field at the expense of the anti-aligned configuration[6] (see also Fig. 3a-c). As previously discussed,[6] the dynamics of this phenomenon is governed by the loss of adhesion energy in the domain wall network and the pinning from the disorder at the interface. The internal out-of-plane coupling reported here (with no external field) reveals a more stable ferroelectric coupling (ABCABC…) in comparison to the antiferroelectric order in the Bernal (ABAB..) configuration.

## S3. Doping and de-polarization measurements

A precise measurement of the out-of-plane polarization at the high doping limit is challenging due to the KPFM signal sensitivity to long-range coulomb forces. The latter interacts with the tip's cantilever and cone rather than its local apex only[7]. While the side-band measurement mode overcomes this challenge to provide quantitative information at zero gate bias, its reliability drops as the external potential on the gate electrode and, correspondingly, the doping charge density on the TMD increase. Crucially, this measurement limitation can only underestimate (by averaging out) the local potential drop, $\Delta V_{KP}$, between domains and its corresponding polarization magnitude. To minimize this underestimation, we used two complementary gating schemes, where either the sample potential is grounded and the gold electrode is biased or vice versa (Fig. S1b, a respectively). Data shown in the main text and in SI are recorded in the configuration of Fig. S1a. We also focused on domains located next to the electrode's edges (while placing the cantilever outside them) and controlled the potential on the global silicon substrate independently. Additional limitations arise from the motion of domain walls at high charge doping and displacement fields, surface chemical adsorption, and surface degradation (see Fig. S2a-c). The latter hinders our quantitative analysis even with a very thin gate dielectric (down to 5 nm thick), where the maximum doping level (at the $h$-BN breakdown electric field) is reached at moderate gate potentials.

Lastly, localized defect states at the host crystals may reduce the occupation of delocalized states by the gate bias. This may result in some overestimation of the precise doping if extracted from the geometric capacitance only. To eliminate this overestimation, we extracted the doping density in Fig. 3d from the change in the average potential ($V_{avg}$), measured on oppositely polarized domains (as marked in Fig. 1b), rather than directly from the applied $V_g$. $V_{avg}$ grows with $V_g$ beyond some threshold value, and in one direction only for each particular sample (Fig. S2d). We attribute this behavior to unpinning of the Fermi-level from gap states associated with native dopants in each sample. The latter seems to prevent achieving electron and hole doping in the same sample. Importantly, $V_{avg}$ is only sensitive to the mobile charge density that accumulates to screen the bottom electrode, regardless of internal properties of the electrodes such as localized defect states, Schottky barriers, or quantum capacitance. The deviation from the ideal $V_{avg} = -V_g$ (dashed black) slope at high $|V_{avg}|$ is attributed to the underestimation of the local KPFM signal in case of large doping levels and spatially alternating potentials at the edges of the electrodes, as discussed above. Altogether, the measurements in Fig. 3d provide an underestimation of the polarization magnitude and the mobile charge density.

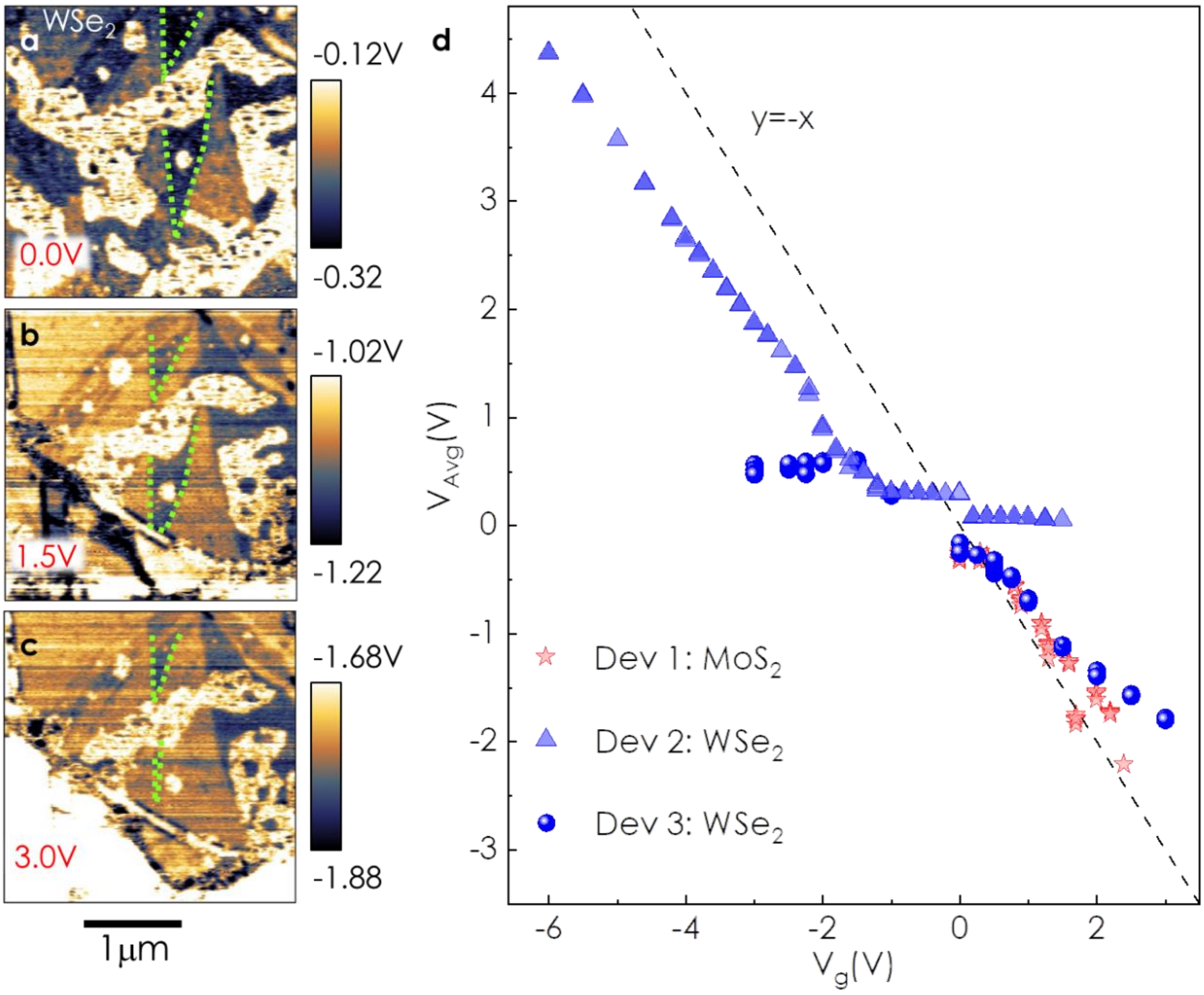

Figure S2. **Doping and depolarization measurements**. (a-c) Examples of surface potential maps obtained from a parallelly stacked $WSe_2$ bilayer under several gate biases $V_g$ (Device 3). Note the domain wall motion at high doping density (dashed green lines) that extends bright areas over dark domains[6]. (d) The average KP potential of the TMD bilayers, $V_{avg}$, as a function of the applied gate potential, $V_g$.

## S4. Calculation of multi-layered $WSe_2$ and $MoS_2$ polarization

In Figs. 1c and 2f of the main text we present the electrostatic potential profile along the normal direction of an ABC stacked $WSe_2$ trilayer and the thickness dependence of the polarization of parallelly stacked $MoS_2$ multilayered systems, respectively. The potential profiels were calculated using the Perdew-Burke-Ernzerhof (PBE) generalized-gradient exchange-correlation density functional approximation,[8] augmented by the Grimme-D3 dispersion correction with Becke-Johnson (BJ) damping[9] as implemented in the Vienna Ab-initio Simulation Package (VASP).[10] The core electrons of the Mo, W, S, and Se atoms were treated via the projector augmented wave (PAW) approach. Spin-orbit interactions were included. This level of theory was recently successfully used to calculate the polarization of transition metal dichalcogenide (TMD) bilayers.[11]

AB stacked $WSe_2$ and $MoS_2$ bilayers, constructed from two relaxed monolayers, were allowed to further relax, yielding lattice constants of 3.29 Å and 3.15 Å, respectively, and interlayer distances (defined as the normal distance between adjacent Se or S ions of the two layers) of 3.10 Å and 2.90 Å, respectively. Single-point electron density calculations were then performed on the relaxed structure with a plane wave energy cutoff of 600 eV and a k-point mesh of $12 \times 12 \times 1$, setting a vertical vacuum size of 10 nm to avoid interactions between adjacent bilayer images. To evaluate the vertical polarization, a dipole moment correction was employed. The potential and charge density profiles along the vertical direction of the two bilayers are shown in Fig. S3. The resulting difference between the electrostatic potential values obtained far above the top and below the bottom surfaces of the $WSe_2$ and $MoS_2$ bilayers are 69 meV and 82 meV, respectively, defining the vertical polarization of the two systems. Multilayered $MoS_2$ systems were constructed on the basis of the AB stacked bilayer by adding additional layers in the AB stacking configuration. Following further optimization, single point potential profile calculations were performed as detailed above.

Convergence tests of the VASP calculations (see Fig. S4) with respect to the vacuum size, energy cut-off, and number of reciprocal space k-points indicate that our choice of parameters leads to $WSe_2$ ($MoS_2$) binding energies that are converged to within 0.006 (0.0001), 0.002 (0.01), and 0.007 (0.003) meV/atom, respectively. Correspondingly, the electrostatic potential difference converges to within 0.01 (0.5), 0.05 (0.001), and 0.04 (0.03) meV with respect to the vacuum size, energy cut-off, and number of reciprocal space k-points, respectively.

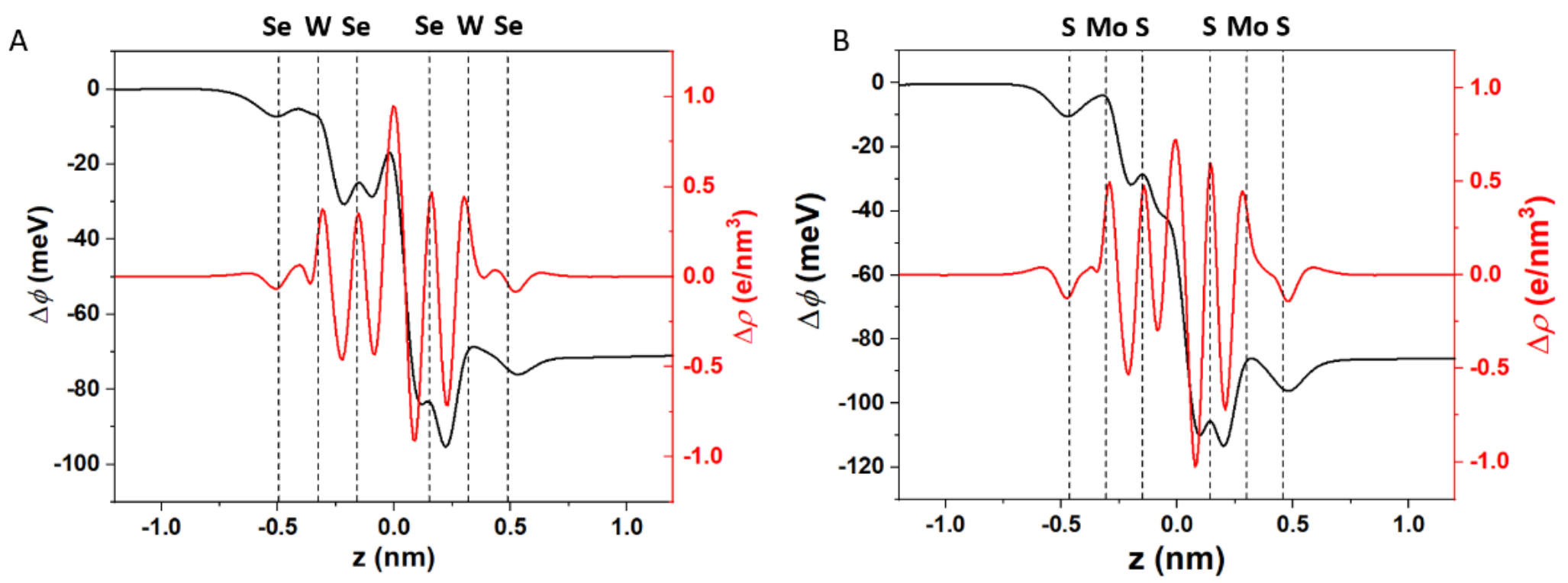

Figure S3. **Potential and charge density profiles of $WSe_2$ and $MoS_2$ bilayers.** Difference between bilayer and isolated monolayer plane-averaged potential (black) and charge density (red) for an AB stacked (a) $WSe_2$ and (b) $MoS_2$ bilayers. The dashed lines represent the vertical location of the ions. The origin of the horizontal axis is set to the midpoint between the layers.

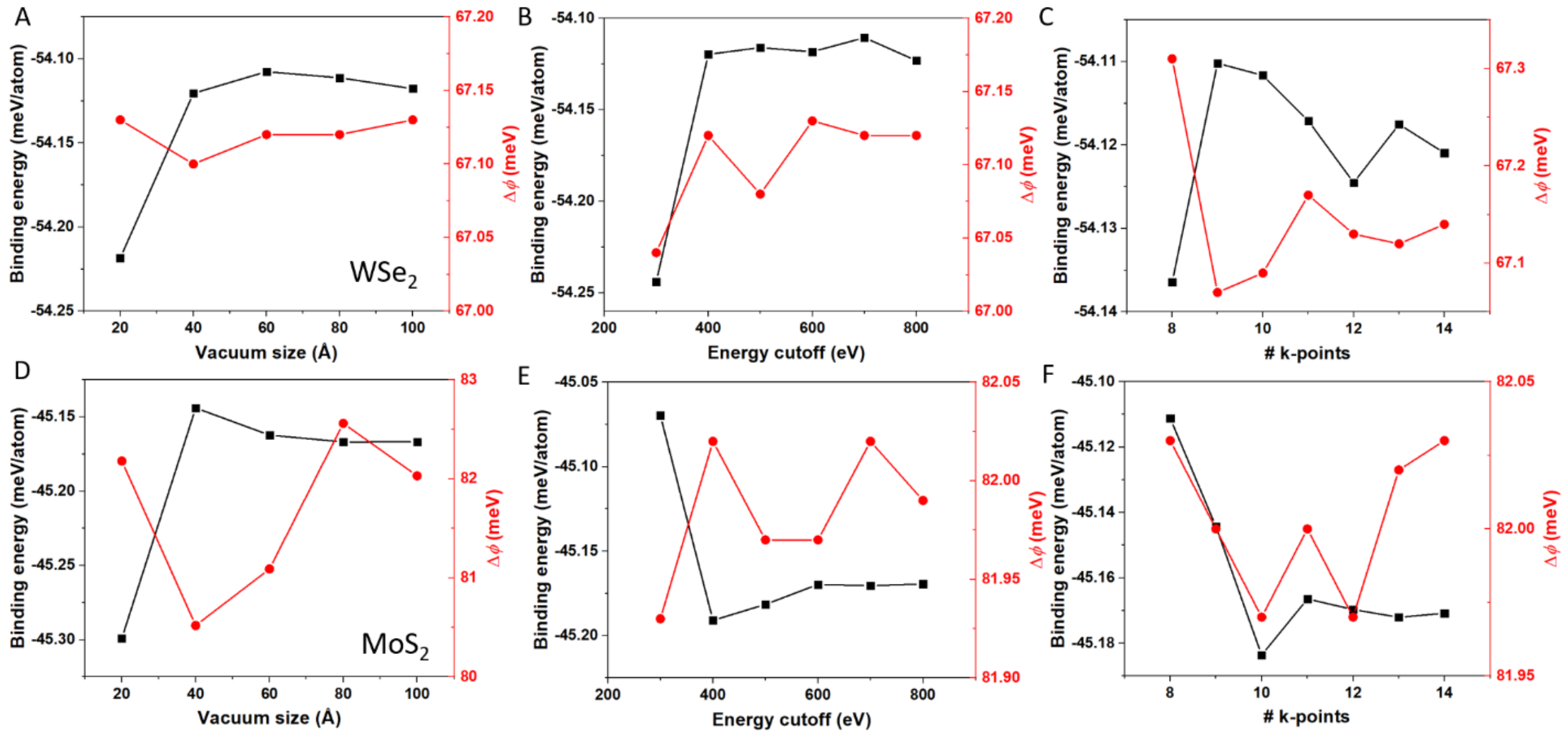

Figure S4. **Convergence tests.** Convergence tests of the binding energy (black curve, left vertical axis) and electrostatic potential difference (red curve, right vertical axis) of $WSe_2$ (top panels) and $MoS_2$ (bottom panels) bilayers (same structures as in Fig. S3) as a function of: (a) vacuum size; (b) energy cutoff, and (c) number of k-points.

## S5. Calculation of doping-induced depolarization in $WSe_2$ and $MoS_2$

Doping calculations of bilayer $WSe_2$ and $MoS_2$ were performed using the fractional nuclear charge pseudoatom approach[12], allowing for simulating doping densities in the experimentally relevant range. To this end, we use pseudopotentials (PPs) generated for atoms with fractional nuclear charge. These calculations were performed using the open source package Quantum Espresso[13], instead of VASP that was used in section S4, allowing us to construct appropriate PPs. We first generated Rappe-Rabe-Kaxiras-Joannopoulos (RRKJ)[14] PPs, including spin-orbit interactions, using the ld1.x program of the plane-wave pseudopotential Quantum Espresso package[13,15]. The nuclear charge of the pseudoatom was set to the original charge of the neutral element plus a small fractional charge $\varepsilon$. For example, the nuclear charge of a doped pseudo W atom was set to $Z = 74 \pm \varepsilon$. The valence electronic charge was changed accordingly to maintain neutrality of the unit-cell, with an electron configuration given by [Xe]$4f^{14}6s^{2}6p^{0}5d^{4\pm\varepsilon}$. A set of PPs were generated by setting $\varepsilon = 10^{-9}, 10^{-8}, \ldots, 10^{-2}$ for all W atoms in the bilayer system, corresponding to doping densities of $\Delta n_{2\mathrm{D}} = 2.1 \times 10^{7}$, $2.1 \times 10^{8}, \ldots, 2.1 \times 10^{13}$ cm$^{-2}$, respectively. A similar procedure was used to generate $MoS_2$ PPs with fractional nuclear charge and valence charge. For example, for a pseudo Mo nuclear charge of $Z = 42 \pm \varepsilon$, the electron configuration was set to [Kr]$5s^{2}5p^{0}4d^{4\pm\varepsilon}$.

Single point calculations were performed using the generated PPs to obtain the electron density and the corresponding electrostatic potential profiles. To this end, we employed the PBE generalized-gradient density functional approximation[8] and the Grimme-D3 dispersion correction with BJ damping,[9] as implemented in Quantum Espresso. A plane wave energy cutoff of 60 Ry (816.34 eV) was used with a k-mesh of $12 \times 12 \times 1$, and a vertical vacuum size of 10 nm was set to avoid interactions between adjacent bilayer images. Fermi-Dirac smearing was used to enhance the convergence of the self-consistent cycle. To obtain the electrostatic potential profiles, a dipole moment correction was used.

As in the procedure discussed in section S4, AB-stacked $WSe_2$ and $MoS_2$ bilayers were first constructed and optimized, yielding lattice constants of 3.29 Å and 3.16 Å and interlayer distances of 3.05 Å and 2.95 Å, respectively. The resulting electrostatic potential drops were 71 meV and 76 meV for the undoped $WSe_2$ and $MoS_2$ bilayers, respectively. Note that little difference (2 and 6 meV for $WSe_2$ and $MoS_2$, respectively) was found between the potential drops calculated by VASP in section S4 and those obtained using Quantum Espresso. The potential and charge density profiles along the vertical direction for the two bilayers are shown in Fig. S5.

Convergence tests for the Quantum Espresso calculations (see Fig. S6) with respect to the vacuum size, energy cut-off, and number of k-points indicate that our choice of parameters leads to $WSe_2$

($MoS_2$) binding energies that are converged to within 0.0003 (0.0003), 0.004 (0.003), and 0.003 (0.0004) meV/atom, respectively. Correspondingly, the electrostatic potential difference converges to within 2.6 (0.9), 3.8 (1.1), and 3.6 (2.4) meV with respect to the vacuum size, energy cut-off, and number of reciprocal space k-points, respectively.

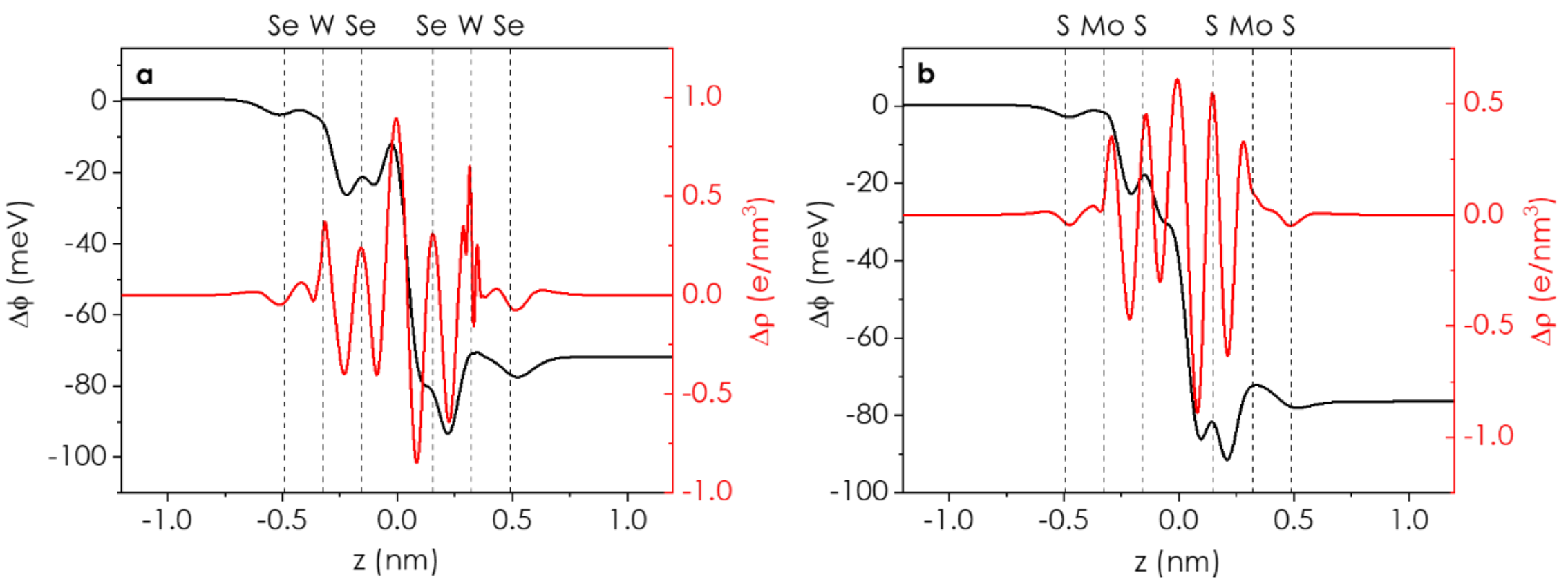

Figure S5. **Potential and charge density profiles for $WSe_2$ and $MoS_2$ bilayer.** Difference between bilayer and isolated monolayer plane-averaged potential (black) and charge density (red) for AB stacked (a) $WSe_2$ and (b) $MoS_2$ bilayers. The dashed lines represent the vertical location of the ions. The origin of the horizontal axis is set to the midpoint between the layers.

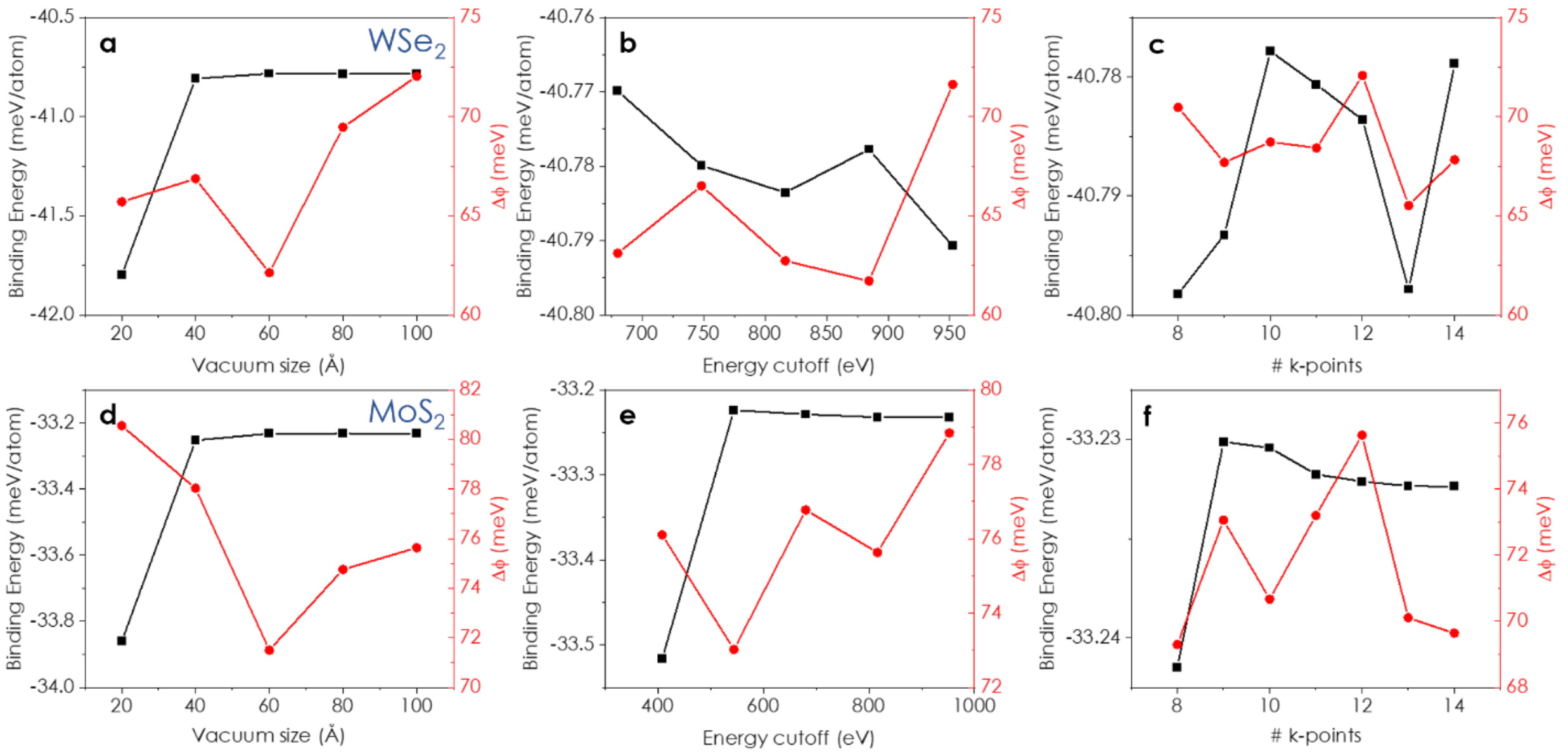

Figure S6. **Convergence tests.** Convergence tests of the binding energy (black curve, left vertical axis) and electrostatic potential difference (red curve, right vertical axis) of $WSe_2$ (top panels) and MoS2 (bottom panels) bilayers (same structures as in Fig. S5) as a function of (a, d) vacuum size; (b, e) energy cutoff, and (c,f) number of k-points.

Doping of the $WSe_2$ and $MoS_2$ bilayers was performed by charging the metal nuclei. As discussed in the main text, up to a system dependent hole or electron charge density, the polarization remains mostly unaffected, following which a polarization drop is clearly seen (see Fig. S7). We note that the fractional nuclear charge pseudoatom doping approach[14] adopted in this study remains valid as long as variations in the calculated band-structure, induced by the nuclear pseudo charging, are negligible. To confirm that our calculations satisfy this condition, we compare the bandstructures of the undoped and doped $WSe_2$ (Fig. S8a) and $MoS_2$ (Fig. S8d) bilayers up to the highest doping density considered. Our results clearly demonstrate merely minor deviations of the band-structures of the doped systems from those of the undoped counterparts. The energy difference between the topmost K and Γ valence band points for the doped and undoped systems is presented in Fig. S8b and S8e for $WSe_2$ and $MoS_2$, respectively. Larger energy differences at higher doping levels result from the depolarization shown in Fig. S7. As an additional validity test, the doping-induced $WSe_2$ and $MoS_2$ Fermi level shifts are presented in Fig. S8c and S8f, respectively, exhibiting the expected logarithmic dependence[16] up to doping densities of $1 \times 10^{13}$ cm$^{-2}$.

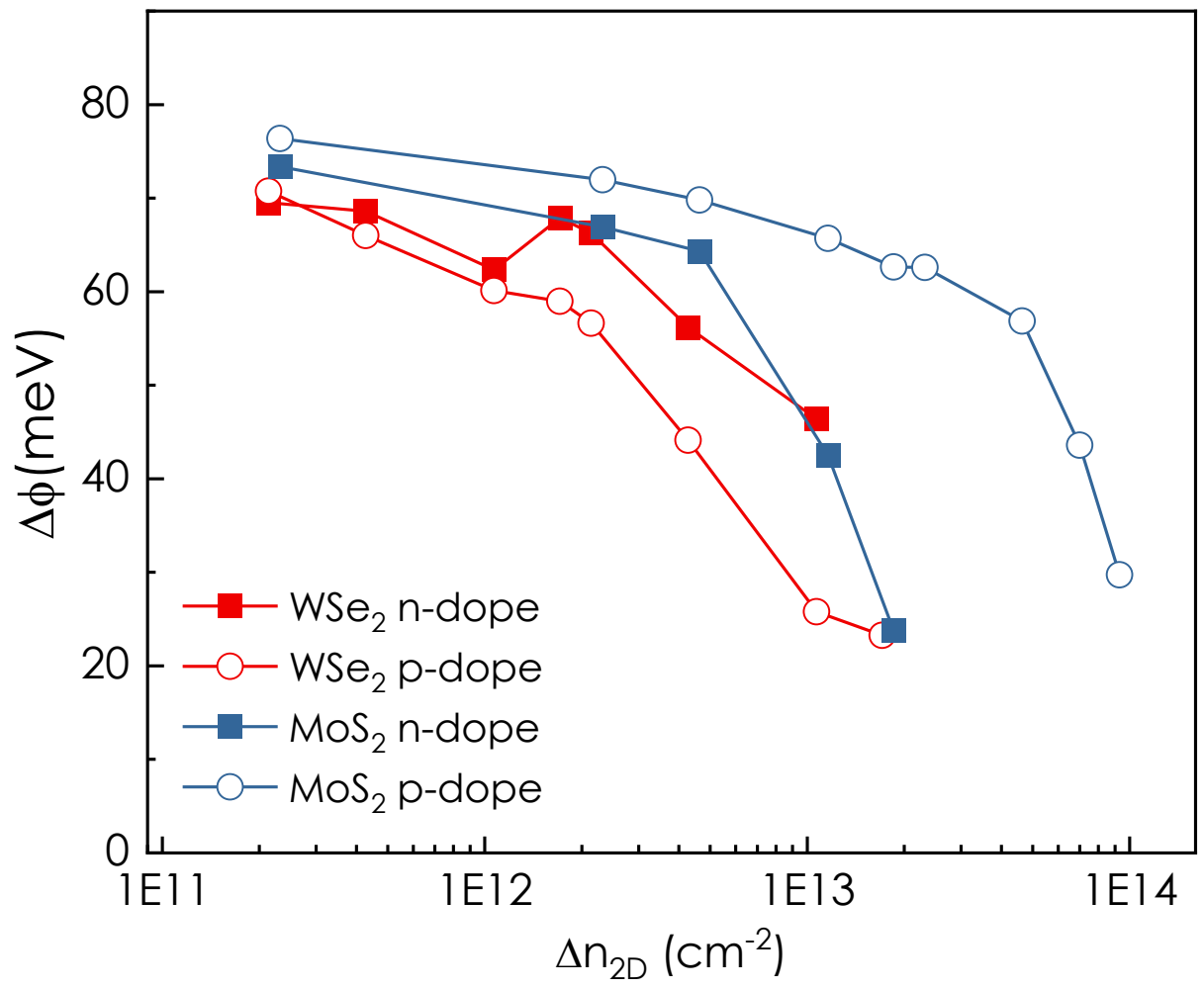

Figure S7. **Effect of doping on the interlayer potential drop.** The potential drop as a function of electron (*n*, filled squares) and hole (*p*, empty circles) doping density for AB stacked bilayer $WSe_2$ (red) and $MoS_2$ (blue). The doping is introduced via the metal pseudo nuclei.

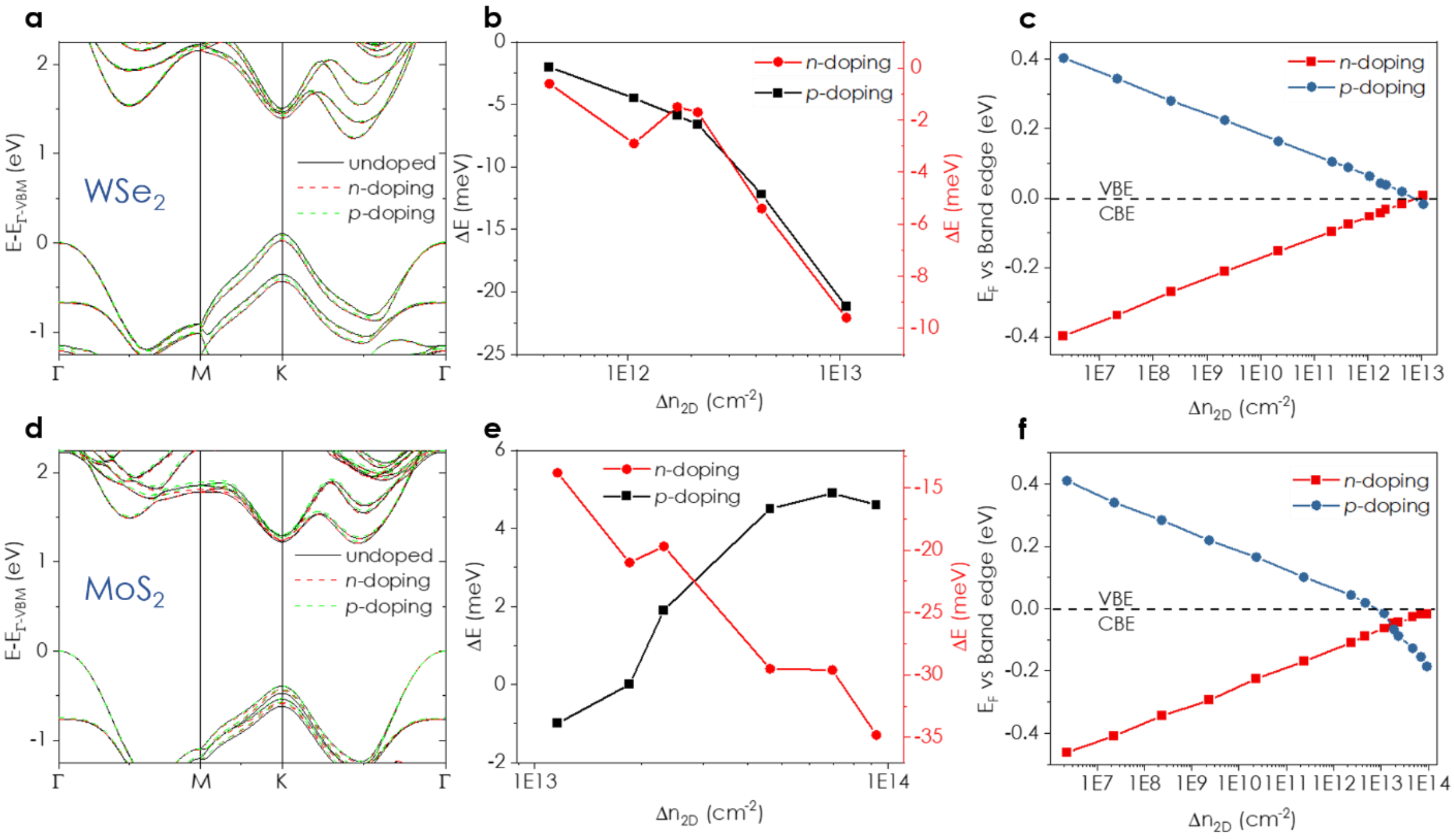

Figure S8. **Band structure and Fermi level variations with doping charge density.** (a, d) The band structures of undoped (black), $n$-doped (red), and $p$-doped (green) $WSe_2$ (a) and $MoS_2$ (d). For $WSe_2$ ($MoS_2$) the $n$-doped and $p$-doped band-structures are plotted for a charge density of $\Delta n_{2D} = \mp 1.1 \times 10^{13}\ cm^{-2}$ ($\mp 9.3 \times 10^{13}\ cm^{-2}$), respectively. The origins of the vertical axes are set to the topmost Γ-point valence band energy ($E_{\Gamma-VBM}$). (b, e) The variation of the difference between the topmost K and Γ valence band energies as a function of doping density for (b) $WSe_2$ and (e) $MoS_2$. Results are presented with respect to the energy difference obtained for the undoped system: $\Delta E = [E_{K-VBM} - E_{\Gamma-VBM}]_{doped} - [E_{K-VBM} - E_{\Gamma-VBM}]_{undoped}$. (c, f) The Fermi level position of (c) $WSe_2$ and (f) $MoS_2$ as a function of $n$-doping (red) and $p$-doping (blue) charge densities. The origins of vertical axes are set to the conduction band minimum energy for $n$-doping and the valence band maximum energy for $p$-doping. The doping is introduced via the metal pseudo nuclei.

To demonstrate that our conclusions are independent on the choice of doping only via the metal atoms, we repeated the calculations by doping only via the chalcogen nuclei or doping all nuclei (see Fig. S9). Consistent results are obtained regardless of the doping scheme.

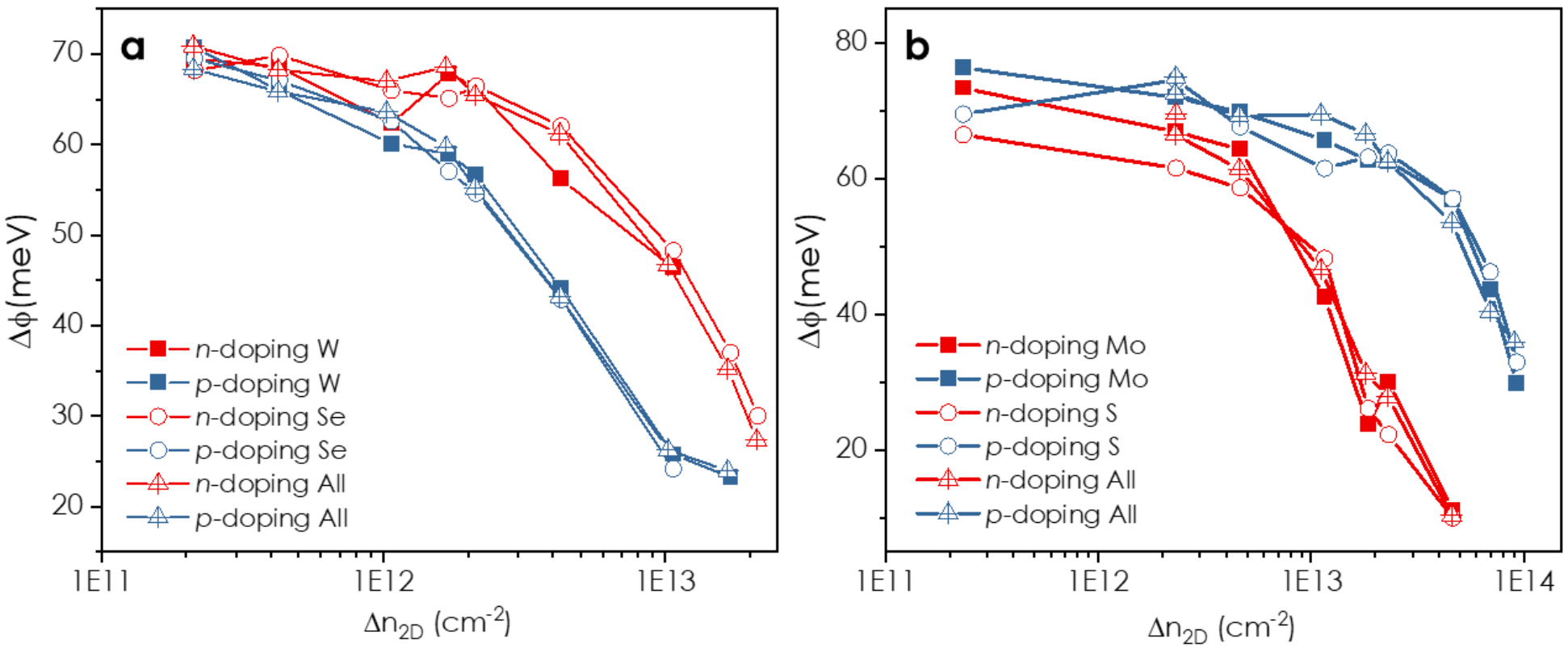

Figure S9. **Comparison of different doping schemes.** The polarization as a function of $n$-doping (red) and $p$-doping (blue) charge density for (a) $WSe_2$ and (b) $MoS_2$. Different doping schemes are applied including: only metal ions (filled squares), only chalcogen ions (open circles), or all ion doping (open triangle). For $WSe_2$, the W ion doping charge excess values were $\varepsilon = 10^{-4}, 2 \times 10^{-4}, 5 \times 10^{-4}, 8 \times 10^{-4}, 10^{-3}, 2 \times 10^{-3}, 5 \times 10^{-3}, 8 \times 10^{-3}, 0.01$; the Se ion doping charge excess values were $\varepsilon = 5 \times 10^{-5}, 10^{-4}, 2.5 \times 10^{-4}, 4 \times 10^{-4}, 5 \times 10^{-4}, 10^{-3}, 2.5 \times 10^{-3}, 4 \times 10^{-3}, 0.005$ ; and the all ion doping charge excess values used were $\varepsilon = 3.3 \times 10^{-5}, 6.6 \times 10^{-5}, 1.6 \times 10^{-4}, 2.6 \times 10^{-4}, 3.3 \times 10^{-4}, 6.6 \times 10^{-4}, 1.6 \times 10^{-3}, 2.6 \times 10^{-3}, 0.0033$. The electron configuration of Se was given by $[\mathrm{Ar}]3d^{10}4s^{2}4p^{4\pm\varepsilon}$, and the corresponding nuclear charge was $Z = 34 \pm \varepsilon$ . For $MoS_2$, the Mo ion doping charge excess values were $\varepsilon = 10^{-4}, 10^{-3}, 2 \times 10^{-3}, 5 \times 10^{-3}, 8 \times 10^{-3}, 0.01, 0.02, 0.03, 0.04$; the S ion doping charge excess values were $\varepsilon = 5 \times 10^{-5}, 5 \times 10^{-4}, 10^{-3}, 2.5 \times 10^{-3}, 4 \times 10^{-4}, 0.005, 0.01, 0.015, 0.02$; and the all ion doping charge excess values used were $\varepsilon = 3.3 \times 10^{-5}, 3.3 \times 10^{-4}, 6.6 \times 10^{-4}, 1.6 \times 10^{-3}, 2.6 \times 10^{-3}, 3.3 \times 10^{-3}, 6.6 \times 10^{-3}, 0.01, 0.013$. The electron configuration of S was given by $[\mathrm{Ne}]3s^{2}3p^{4\pm\varepsilon}$, and the corresponding nuclear charge was $Z = 16 \pm \varepsilon$.

# S6. Effect of polarization on the band-structure

To evaluate the effect of the emerging polarization on the band structure, we compare in Fig. S10 the band structure of the anti-parallel AA' stacked undoped bilayers with those of the parallel AB stacked counterparts, all evaluated at the same level of theory as described in section S5. The results clearly demonstrate band splitting of both the conduction and the valence bands at the K point. Notably, this splitting is of the order of the calculated vertical potential drops (see SI section S5 above), indicating that the emerging polarization is indeed causing the splitting.

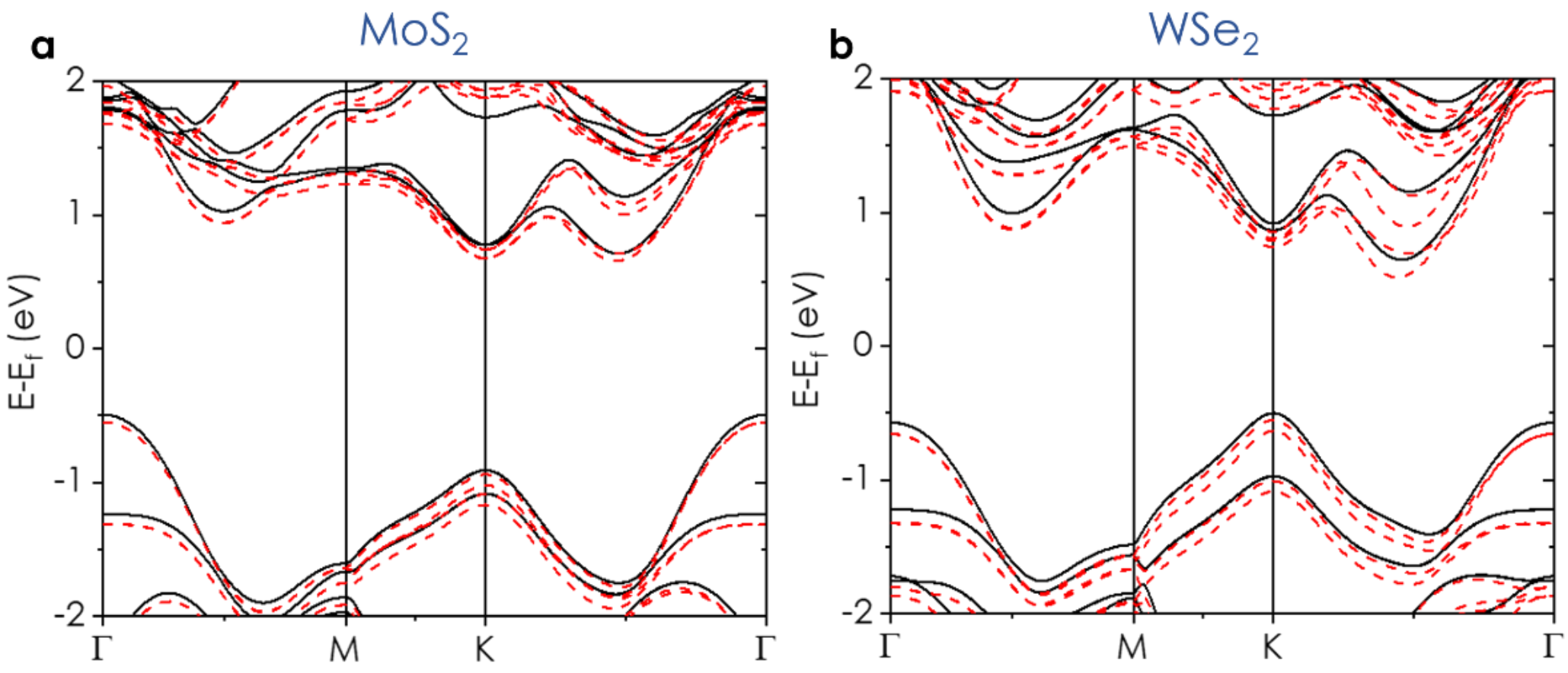

Figure S10. **Effect of polarization on the band structure.** The band structures of (a) $MoS_2$ and (b) $WSe_2$ bilayers at their anti-parallel AA' (solid black curve) and parallel AB (dashed red curve) stacking modes.